\documentclass[twocolumn,showpacs,preprintnumbers,amsmath,amssymb,APSl,prd,nofootinbib,superscriptaddress]{revtex4-1}

\usepackage{dcolumn}
\usepackage{bm}
\usepackage{ifpdf}
\usepackage{hyperref}
\usepackage{dcolumn}
\usepackage{bm}
\usepackage[spanish,english]{babel}
\usepackage{amsfonts}
\usepackage{amssymb}
\usepackage{graphicx}
\usepackage[latin1]{inputenc}

\newcommand{\LL}{\mathcal{L}}

\newcommand{\be}{\begin{equation}}
\newcommand{\en}{\end{equation}}
\newcommand{\bea}{\begin{eqnarray}}
\newcommand{\ena}{\end{eqnarray}}

\begin{document}

\title{Born-Infeld gravity and its functional extensions}

\author{Sergei D. Odintsov} 
\affiliation{Instituci\`{o} Catalana de Recerca i Estudis Avan\c{c}ats (ICREA), Barcelona, 08010 Spain}
\affiliation{Institut de Ciencies de l'Espai (CSIC-IEEC), Campus UAB,
 Torre C5-Par-2a- pl, E-08193 Bellaterra (Barcelona), Spain}
\affiliation{King Abdulaziz University, Jeddah, 21441 Saudi Arabia}
\affiliation{Tomsk State Pedagogical University, Kievskaya Street 60, 634061 Tomsk, Russia}
\author{Gonzalo J. Olmo} \email{gonzalo.olmo@csic.es}
\affiliation{Departamento de F\'{i}sica Te\'{o}rica and IFIC, Centro Mixto Universidad de
Valencia - CSIC. Universidad de Valencia, Burjassot-46100, Valencia, Spain}
\affiliation{Departamento de F\'isica, Universidade Federal da
Para\'\i ba, 58051-900 Jo\~ao Pessoa, Para\'\i ba, Brazil}
\author{D. Rubiera-Garcia} \email{drubiera@fisica.ufpb.br}
\affiliation{Departamento de F\'isica, Universidade Federal da
Para\'\i ba, 58051-900 Jo\~ao Pessoa, Para\'\i ba, Brazil}


\date{\today}

\begin{abstract}
We investigate the dynamics of a family of functional extensions of the (Eddington-inspired) Born-Infeld gravity theory, constructed with the inverse of the metric and the Ricci tensor. We provide a generic formal solution for the connection and an Einstein-like representation for the metric field equations of this family of theories. For particular cases we consider applications to the early-time cosmology and find that non-singular universes with a cosmic bounce are very generic and robust solutions.
\end{abstract}


\maketitle

\section{Introduction}

The success of the current standard model for cosmology including an early phase of inflationary expansion cannot be underestimated. It provides a description of the cosmic evolution over a period of more than $13$ billion years that is in excellent agreement with a number of high-precision and independent observations. Nevertheless, despite its observational success, the model still leaves without a fully satisfactory answer a number of open questions, such as the conditions that originated the inflationary process.  This is the epoch at which quantum gravitational effects are expected to manifest themselves and solve issues such as the existence of a classical big bang singularity.

Different approaches directly dealing with the quantization of gravity have shed interesting light on the pre-inflationary cosmic dynamics in the last years  \cite{Ashtekar:2011ni,McAllister:2007bg}. A rather generic prediction is the existence of an earlier phase of cosmic contraction followed by a bounce that sets the beginning of our current expansion phase \cite{Novello}. The minimum size attained by the Universe in this process would replace the  classically predicted, zero-volume big bang singularity. In these scenarios, therefore, an effective geometry with modified dynamics arises and avoids the shortcomings of unbounded curvature scalars of classical General Relativity (GR).

Though quantum-motivated cosmological models have shown their ability to get rid of the big bang and other singularities, one might wonder if the disturbing aspects of singularities could be avoidable by means of classical improvements of the theory. A source of inspiration in this sense can be found in the Born-Infeld (BI) theory of classical electrodynamics \cite{BIem}, where the field strength and the self-energy of the electron become bounded, thus avoiding those divergences found in the standard Maxwell theory. A Born-Infeld type action for the gravitational field, dubbed Eddington-inspired Born-Infeld gravity (Born-Infeld theory in what follows, for simplicity), has been studied recently \cite{Deser:1998rj,Banados} with interesting results in cosmology and black hole physics (see \cite{BI-extensions1}-\cite{BI-extensions10} for further proposal on Born-Infeld-type gravities). It has been found that the theory is able to avoid the big bang singularity yielding a cosmic bounce in simple scenarios involving a radiation fluid or even a pressureless dust fluid \cite{Banados}. Certain configurations of electrically charged black holes are also able to avoid the central singularity, which is generically replaced by a smooth wormhole \cite{Olmo:2013gqa}. The implications of this theory have been thoroughly investigated in cosmology \cite{Du:2014jka, Kim:2013noa,Kruglov:2013qaa, Yang:2013hsa,Avelino:2012ue, DeFelice:2012hq, EscamillaRivera:2012vz, Cho:2012vg, Scargill:2012kg, EscamillaRivera:2013hv}, astrophysics \cite{Harko:2013xma,Avelino:2012ge}, stellar structure \cite{Sham:2013cya,Kim:2013nna,Harko:2013wka,Sham:2013sya, Avelino:2012qe, Sham:2012qi,Pani:2012qd,  Pani:2011mg}, the problem of cosmic singularities \cite{Bouhmadi-Lopez:2013lha, Ferraro:2010at}, black holes \cite{Olmo:2013gqa}, and wormhole physics \cite{Lobo:2014fma,Harko:2013aya}, among many others.

In an attempt to explore the robustness of the predictions of the BI theory, a program considering a family of extensions of this theory was initiated in \cite{Makarenko:2014lxa,Makarenko:2014nca}, where an $f(R)$ piece was added to the original theory as a way to consider new curvature interactions or tune the parameters of the $R$-dependent terms that appear in a series expansion of the theory. It was found that cosmological  bouncing solutions are robust against $R^2$ modifications of the Lagrangian. Additionally, solutions that were non-singular but unstable in the original theory, could develop a big bang instability followed by a phase of cosmological inflation ($H\sim$ constant) even in radiation dominated scenarios, with a smooth transition to a standard expansion in good agreement with general relativity.

Here we show that a different and simpler extension of the BI theory exists and explore its cosmological implications at early times. We find that BI gravity can be naturally written as a kind of $f(|\hat\Omega|)$ theory where $f(|\hat\Omega|)=|\hat\Omega|^{1/2}$ and $|\hat\Omega|$ is the determinant of a matrix $\hat\Omega$ defined as $\hat\Omega=\hat g^{-1} \hat q$, with $\hat q$ denoting the matrix representation of $q_{\mu\nu}\equiv g_{\mu\nu}+\epsilon R_{\mu\nu}$. Here $g_{\mu\nu}$ is the space-time metric and $R_{\mu\nu}$ is the Ricci tensor associated to the connection, which is assumed to be independent of the metric {\it a priori} (Palatini formalism).
We provide a detailed derivation of the action and field equations, and then consider applications to cosmology. By constructing parametric representations of the Hubble function as a function of the matter density, we explore the impact on the solutions of modifying the form of the Lagrangian. In particular, we consider extensions of the form $f(|\hat\Omega|)=|\hat\Omega|^{n}$, as they allow a straightforward treatment of the field equations. Numerical methods are, however, necessary in general to carry out the analysis of particular models and equations of state. This type of models modify the cosmic dynamics at high energies, leaving the low-energy dynamics essentially unchanged as compared to GR. From an observational point of view, therefore, these extensions constitute viable theories.
From the theoretical side, they offer a new way to explore the dynamics of the early universe and the robustness of some basic predictions.

\section{Born-Infeld theory and the $\hat\Omega$ representation}

The standard form for the action of Born-Infeld gravity is written as follows
\begin{eqnarray}
S_{BI}&=&\frac{1}{\kappa^2\epsilon}\int d^4x \left[\sqrt{-|g_{\mu\nu}+\epsilon R_{\mu\nu}(\Gamma)|}-\lambda \sqrt{-|g_{\mu\nu}|}\right] \\
&+&S_m(g_{\mu\nu},\psi_m), \nonumber
\end{eqnarray}
where $\kappa^2$ is a constant with suitable dimensions (in GR, $\kappa^2=8\pi G/c^3$),  $|g_{\mu\nu}|$ is the determinant of the space-time metric $g_{\mu\nu}$, $\epsilon$ is a parameter with dimensions of a squared length, $R_{\mu\nu}(\Gamma)$ is the Ricci tensor constructed with the affine connection $\Gamma \equiv \Gamma_{\mu\nu}^{\lambda}$, which is a priori independent of the metric structure (Palatini formalism), $S_m$ is the matter action and $\psi_m$ denote collectively the matter fields, which only couple to the metric as dictated by the equivalence principle. The parameter $\lambda$ is of order $\sim 1$ and its meaning will become clear later. In order to find functional extensions of this theory, we should write it in the form $S=\int d^4x \sqrt{-g}\LL_G$. This can be done by noting that the first term in the integrand can be written as
\begin{equation}
\sqrt{-|g_{\mu\nu}+\epsilon R_{\mu\nu}(\Gamma)|}= \sqrt{-|g_{\mu\alpha}\left({\delta^\alpha}_{\nu}+\epsilon g^{\alpha\beta}R_{\beta\nu}(\Gamma)\right)|} \ .
\end{equation}
This decomposition simply states the fact that the matrix $q_{\mu\nu}\equiv g_{\mu\nu}+\epsilon R_{\mu\nu}(\Gamma)$ can be written as $q_{\mu\nu}=g_{\mu\alpha}\left({\delta^\alpha}_\nu+\epsilon{P^\alpha}_\nu\right)$, where ${P^\alpha}_\nu\equiv g^{\alpha\beta}R_{\beta\nu}(\Gamma)$. From now on, we will denote $q_{\mu\nu}=g_{\mu\alpha}{\Omega^\alpha}_\nu$, where ${\Omega^\alpha}_\nu\equiv g^{\alpha\beta}q_{\beta\nu}= {\delta^\alpha}_\nu+\epsilon{P^\alpha}_\nu$. In matrix notation, we have $\hat\Omega={\hat g^{-1}}\hat q$ and $\hat\Omega^{-1}={\hat q^{-1}}\hat g$. With this notation, the Born-Infeld action becomes
\begin{equation}
S_{BI}=\frac{1}{\kappa^2\epsilon}\int d^4x \sqrt{-g}\left[\sqrt{|\hat\Omega|}-\lambda \right]+S_m \ .
\end{equation}
This representation suggests the following family of theories:
\begin{equation}\label{eq:f(BI)a}
S_f=\frac{1}{\kappa^2\epsilon}\int d^4x \sqrt{-g}\left[f(|\hat\Omega|)-\lambda \right]+S_m \ ,
\end{equation}
being $f(|\hat\Omega|)=|\hat\Omega|^{1/2}$ in the Born-Infeld case.

\section{Field equations of $f(|\hat\Omega|)$ theories}

In order to obtain the field equations corresponding to the theory (\ref{eq:f(BI)a}), we first re-write that action introducing an auxiliary scalar field $A$ such that \cite{Olmo:2005zr}
\begin{equation}\label{eq:f(BI)b}
S_f=\frac{1}{\kappa^2\epsilon}\int d^4x \sqrt{-g}\left[f(A)+(|\hat\Omega|-A)f_A-\lambda \right]+S_m \ ,
\end{equation}
which can be expressed in Brans-Dicke form as
\begin{equation}\label{eq:f(BI)c}
S_f=\frac{1}{\kappa^2\epsilon}\int d^4x \sqrt{-g}\left[\phi|\hat\Omega|-V(\phi)-\lambda \right]+S_m \ ,
\end{equation}
where $\phi\equiv df/dA$ and $V(\phi)=A(\phi) f_A-f(A)$. Note that from the definition $\phi\equiv df/dA$ one can obtain an expression for $A=A(\phi)$, which is necessary to construct $V(\phi)$. Variation of $S_f$ in this last representation, yields
\begin{eqnarray}
\delta S_f&=&\frac{1}{\kappa^2\epsilon}\int d^4x \sqrt{-g}\Big[\frac{\epsilon\kappa^2\LL_G}{2}g^{\mu\nu}\delta g_{\mu\nu} \nonumber \\ &+&\Big(|\hat\Omega|-V_\phi\Big)\delta\phi + \phi \delta |\hat\Omega| \Big]+\delta S_m \ ,
\end{eqnarray}
where
\begin{equation} \label{eq:Lagrangian}
\epsilon\kappa^2\LL_G=\phi|\hat\Omega|-V(\phi)-\lambda.
\end{equation}
A key step now is to note that ${\Omega^\alpha}_\nu\equiv g^{\alpha\beta}q_{\beta\nu}$ and that $\delta |\hat\Omega|=|\hat\Omega| {[\hat{\Omega}^{-1}]^\nu}_\alpha \delta  {\Omega^\alpha}_\nu$, where
\begin{eqnarray}
\delta {\Omega^\alpha}_\nu&=&q_{\beta\nu} \delta g^{\alpha\beta}+g^{\alpha\beta}  \delta q_{\beta\nu} \nonumber \\
&=& -g^{\alpha\sigma}g^{\beta\rho}q_{\beta\nu} \delta g_{\sigma\rho}+g^{\alpha\beta}  \delta q_{\beta\nu}\nonumber \\
&=& -g^{\alpha\sigma} {\Omega^\rho}_\nu \delta g_{\sigma\rho}+g^{\alpha\beta}  \left(\delta g_{\beta\nu}+\epsilon \delta R_{\beta\nu}(\Gamma)\right) \ .
\end{eqnarray}
This leads to
\begin{eqnarray}
\delta |\hat\Omega|&=&|\hat\Omega| {[\hat{\Omega}^{-1}]^\nu}_\alpha \Big[-g^{\alpha\sigma}{\Omega^\rho}_\nu  \delta g_{\sigma\rho} \nonumber \\
&+& g^{\alpha\beta}  \Big(\delta g_{\beta\nu}+\epsilon \delta R_{\beta\nu}(\Gamma)\Big) \Big] \\
&=&|\hat\Omega|  \left[-g^{\mu\nu} \delta g_{\mu\nu}+ [\hat{q}^{-1}]^{\mu\nu} \left(\delta g_{\mu\nu}+\epsilon \delta R_{\mu\nu}(\Gamma)\right) \right] \nonumber ,
\end{eqnarray}
and inserting this result in $\delta S_f$, we get
\begin{eqnarray}
\delta S_f&=&\frac{1}{\kappa^2\epsilon}\int d^4x \sqrt{-g}\Big[\Big(\frac{\epsilon\kappa^2\LL_G}{2}g^{\mu\nu} \nonumber \\
&+& \phi|\hat\Omega| (-g^{\mu\nu}+ [\hat{q}^{-1}]^{\mu\nu} )\Big)\delta g_{\mu\nu} \\ &+&\left.\left(|\hat\Omega|-V_\phi\right)\delta\phi+\epsilon \phi |\hat\Omega| [\hat{q}^{-1}]^{\mu\nu}\delta R_{\mu\nu}(\Gamma)\right]+\delta S_m \ .\nonumber
\end{eqnarray}
The variation of the Ricci tensor (discarding all torsional terms at the end of the calculations for simplicity \cite{Olmo:2013lta}), $\delta R_{\mu\nu}=\nabla_\lambda \delta\Gamma^\lambda_{\nu\mu}-\nabla_\nu \delta\Gamma^\lambda_{\lambda\mu}$, can be integrated by parts leading to
\begin{eqnarray}
\delta S_f&=&\frac{1}{\kappa^2\epsilon}\int d^4x \sqrt{-g}\Big[\Big(\frac{\epsilon\kappa^2 \LL_G}{2}g^{\mu\nu} \nonumber \\
&+&\phi|\hat\Omega| (-g^{\mu\nu}+ {q}^{\mu\nu} )\Big)\delta g_{\mu\nu}+\Big(|\hat\Omega|-V_\phi\Big)\delta\phi \nonumber \\ &-&\frac{\epsilon}{\sqrt{-g}} \Big(\nabla_\lambda\Big[\phi \sqrt{-g} |\hat\Omega| {q}^{\mu\nu} \Big] \\
&-& {\delta^\nu}_\lambda \nabla_\sigma\Big[\phi \sqrt{-g} |\hat\Omega| {q}^{\mu\sigma}\Big]\Big) \delta\Gamma^\lambda_{\nu\mu} \Big]+\delta S_m, \nonumber
\end{eqnarray}
where we have simplified the notation denoting $[\hat{q}^{-1}]^{\mu\nu}\equiv {q}^{\mu\nu}$ to represent the inverse of the tensor $q_{\mu\nu}$. It is important to note that $ {q}^{\mu\nu}\neq g^{\mu\alpha}g^{\nu\beta}q_{\alpha\beta}$. The field equations can thus be written as follows:
\begin{eqnarray}\label{eq:gvar}
\phi |\hat\Omega| {q}^{\mu\nu}-\frac{\left(\phi |\hat\Omega|+V(\phi)+\lambda\right)}{2}g^{\mu\nu}&=&-\frac{\kappa^2\epsilon}{2}T^{\mu\nu} \\
\left(|\hat\Omega|-\frac{dV}{d\phi}\right)&=&0 \label{eq:V_phi}\\
\nabla_\lambda\left[\phi \sqrt{-g} |\hat\Omega| {q}^{\mu\nu}\right]&=&0. \label{eq:connection1}
\end{eqnarray}
The second of these field equations simply establishes an algebraic relation between the potential $V(\phi)$ and the determinant $|\hat\Omega|$, which allows to express $\phi$ as a function of $|\hat\Omega|$. To solve the other two equations we note that, without knowing the particular $f(|\hat\Omega|)$ theory employed, the connection equation can be solved, in general, using algebraic methods \cite{OSAT}. To see this, recall that $\hat\Omega={\hat g}^{-1}\hat q$, which implies that $|\hat\Omega|=|q|/|g|$ and, therefore, $\sqrt{-g}=\sqrt{-q}\Omega^{-1/2}$. As a result, (\ref{eq:connection1}) becomes
\begin{equation}\label{eq:connection2}
\nabla_\lambda\left[\phi |\hat\Omega|^{1/2}\sqrt{-q}  {q}^{\mu\nu}\right]=0 \ .
\end{equation}
Introducing an auxiliary tensor
\begin{equation}
t_{\mu\nu}=\phi |\hat\Omega|^{1/2} q_{\mu\nu},
\end{equation}
we find that  $t^{\mu\nu}= q^{\mu\nu}/(\phi |\hat\Omega|^{1/2})$ and $|t|^{1/2}=\phi^2  |\hat\Omega|q|^{1/2}$, which turns  (\ref{eq:connection2}) into
\begin{equation}\label{eq:connection3}
\nabla_\lambda\left[\sqrt{-t}  {t}^{\mu\nu}\right]=0 \ .
\end{equation}
This equation implies that, regardless of the theory chosen, the independent connection is the Levi-Civita connection of the auxiliary metric $t_{\mu\nu}=\phi |\hat\Omega|^{1/2} q_{\mu\nu}$. The reason for this manipulation lies in the fact that, as will be clear shortly, the field equations in terms of the metric $t_{\mu\nu}$ can be cast in an Einstein-like form, which facilitates their investigation in physical applications.

Before getting into those aspects, let us first clarify the meaning and dependences of $\phi$ and $|\hat\Omega|$. Note that (\ref{eq:V_phi}) determines the explicit relation between $\phi$ and $|\hat\Omega|$, which is an algebraic equation and implies $\phi=\phi(|\hat\Omega|)$. On the other hand, using the definition $\hat\Omega^{-1}={\hat q}^{-1}\hat g$ in (\ref{eq:gvar}), one finds
\begin{equation}\label{eq:Omega-T}
\phi |\hat\Omega| {[{\hat\Omega}^{-1}]^\mu}_\nu=\frac{\left(\phi |\hat\Omega|+V(\phi)+\lambda\right)}{2}{\delta^{\mu}}_\nu-\frac{\kappa^2\epsilon}{2}{T^{\mu}}_\nu \ .
\end{equation}
Since $\phi$ is a function of $|\hat\Omega|$, (\ref{eq:Omega-T}) establishes an algebraic relation between the matrix ${[{\hat\Omega}^{-1}]^\mu}_\nu$ and the matter stress-energy tensor ${T^{\mu}}_\nu$. With the knowledge of ${[{\hat\Omega}^{-1}]^\mu}_\nu$ one can easily obtain an expression for $q_{\mu\nu}$ that only involves the metric $g_{\mu\nu}$ and the matter. As a result, Eq. (\ref{eq:connection3}) needs not be seen as an equation involving up to second-order derivatives of
the connection $\Gamma$, rather it is an equation that depends linearly on $\Gamma$, on the first derivatives of the metric $g_{\mu\nu}$ and on derivatives of $|\hat\Omega|$ (which is a function of the matter stress-energy tensor). The connection $\Gamma$ is, therefore, a non-dynamical object completely determined by $g_{\mu\nu}$ and the matter.

Summarizing, equations (\ref{eq:gvar}), (\ref{eq:V_phi}), and (\ref{eq:connection1}) establish a set of {\it algebraic} relations among various quantities of the theory. In the next section we shall obtain the dynamical field equations for the metric.

\section{Metric field equations}

Now we derive the form of the differential equations that govern our theory. To do it, we begin with the definition $q_{\mu\nu}=g_{\mu\nu}+\epsilon R_{\mu\nu}(\Gamma)$,  and multiply it by the inverse of $q_{\mu\nu}$ to obtain
$\epsilon q^{\mu\alpha} R_{\alpha\nu}(t)={\delta^{\mu}}_\nu-{ [\hat\Omega^{-1}]^{\mu}}_\nu$, where we have used the notation $R_{\mu\nu}(\Gamma)=R_{\mu\nu}(t)$ because the connection $\Gamma$ is the Levi-Civita one of $t_{\mu\nu}$. Since $t^{\mu\alpha}=q^{\mu\alpha} /(\phi|\hat\Omega|^{1/2})$, we have
\begin{equation}
{R^\mu}_\nu(t)\equiv t^{\mu\alpha}R_{\alpha\nu}(t)=\frac{1}{\epsilon \phi|\hat\Omega|^{1/2}}\left({\delta^{\mu}}_\nu-{ [\hat\Omega^{-1}]^{\mu}}_\nu\right)
\end{equation}
Using now Eq.(\ref{eq:Omega-T}), we get
\begin{equation}\label{eq:Rmn-t}
{R^\mu}_\nu(t)=\frac{\kappa^2}{2 \phi^2|\hat\Omega|^{3/2}}\left(\LL_G{\delta^{\mu}}_\nu+{T^{\mu}}_\nu\right) \ ,
\end{equation}
where $\LL_G\equiv (\phi|\hat\Omega|-V(\phi)-\lambda)/(\epsilon\kappa^2)$. In general, both $\phi$ and $|\hat\Omega|$ appearing on the right-hand side of  (\ref{eq:Rmn-t}) are functions of the matter, and therefore, all the right-hand side is a function of the matter sources. We can thus solve for $t_{\mu\nu}$ and then use the transformation law $t_{\mu\nu}=\phi |\hat\Omega|^{1/2}g_{\mu\alpha}{\Omega^{\alpha}}_\nu $ to obtain $g_{\mu\nu}=t_{\mu\alpha}{[\hat\Omega^{-1}]^{\alpha}}_\nu/(\phi |\hat\Omega|^{1/2})$, which provides a complete solution for a given Lagrangian $f(|\hat\Omega|)$ and energy-momentum tensor $T_{\mu\nu}$. It should be stressed that in vacuum, ${T_\mu}^{\nu}=0$, due to the dependence of $\phi$ and ${\Omega^\alpha}_{\nu}$ in ${T_\mu}^{\nu}$ the field equations boil down to those of GR with a cosmological constant term.

This concludes the formulation and analysis of this type of $f(|\hat\Omega|)$ theories. Applications are now possible.

\section{Perfect fluids}

\begin{figure}[h]
\begin{center}
\includegraphics[width=0.5\textwidth]{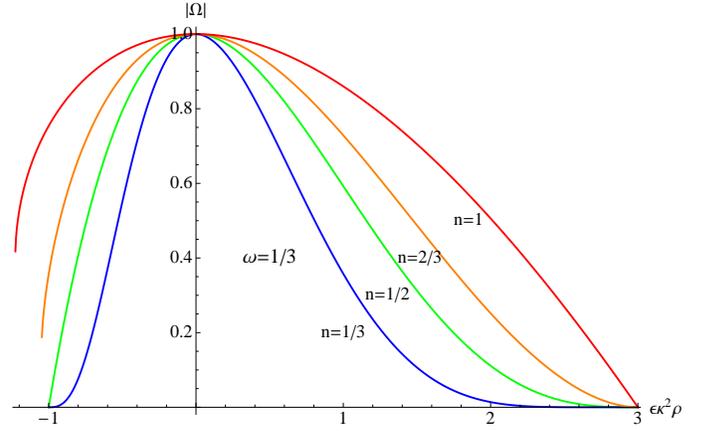}
\caption{Representation of the function $|\hat\Omega|(\rho)$ for a radiation fluid, $\omega=1/3$, in a family of theories defined as $f(|\hat\Omega|)=|\hat\Omega|^n$ (recall that the Born-Infeld theory corresponds to $n=1/2$) and with $\lambda=1$. The negative horizontal axis corresponds to models with $\epsilon<0$. The curves cover all the physical domain of the variable $\epsilon \kappa^2\rho$. \label{fig:w1b3}}
\end{center}
\end{figure}

\begin{figure}[h]
\begin{center}
\includegraphics[width=0.5\textwidth]{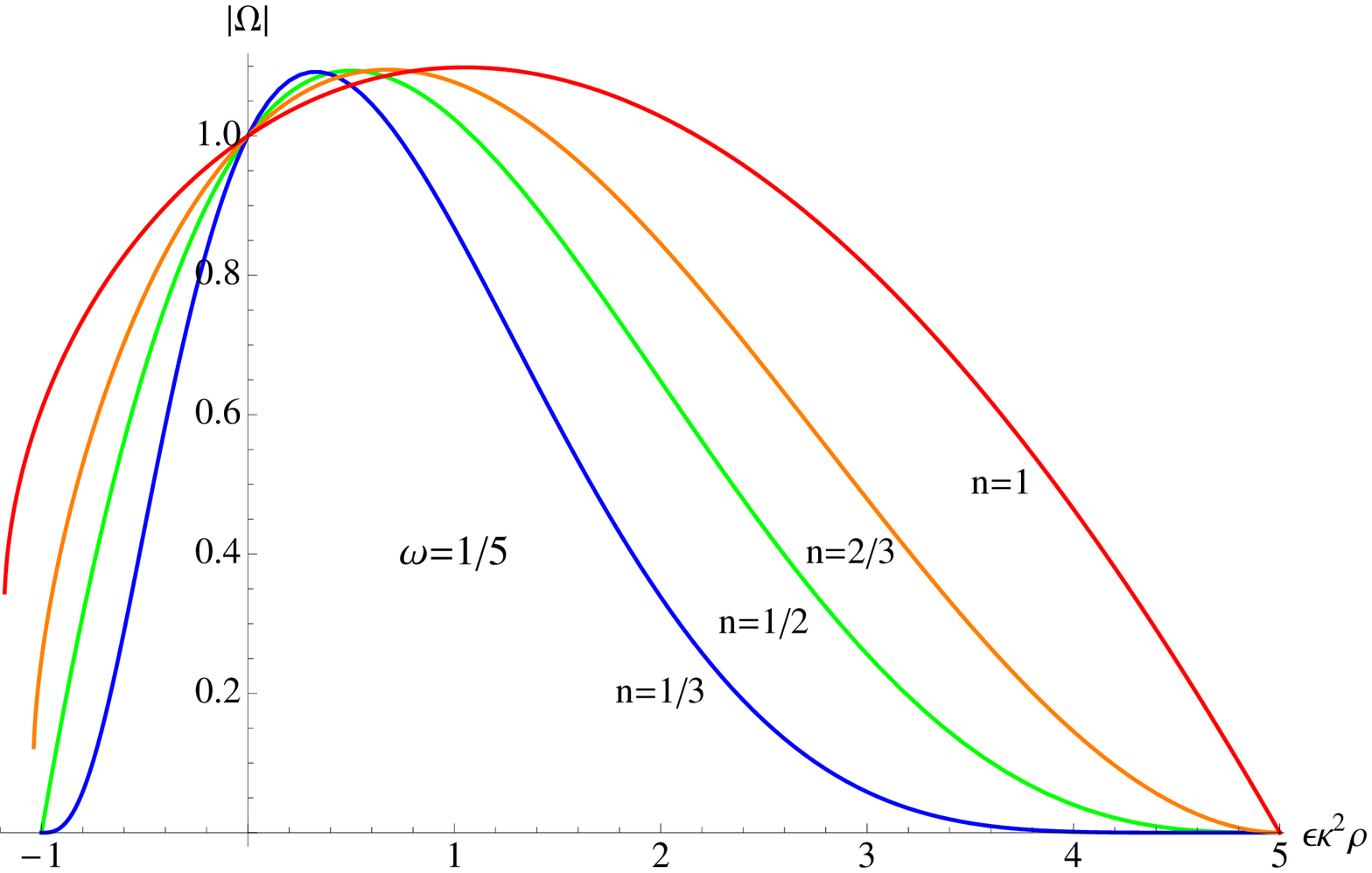}
\caption{Same representation as in Fig. \ref{fig:w1b3} but for the case  $\omega=1/5$. \label{fig:w1b5}}
\end{center}
\end{figure}

\begin{figure}[h]
\begin{center}
\includegraphics[width=0.5\textwidth]{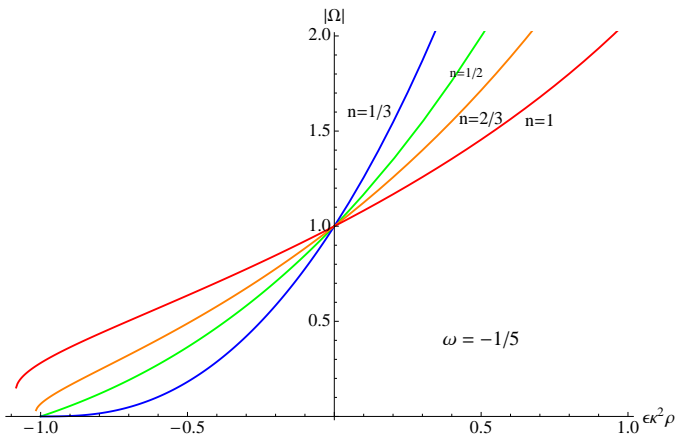}
\caption{Same representation as in Fig. \ref{fig:w1b3} but for the case  $\omega=-1/5$. The curves on the first quadrant are divergent. \label{fig:wm1b5}}
\end{center}
\end{figure}

For a perfect fluid with energy-momentum tensor ${T^{\mu}}_\nu=(\rho+P)u^\mu u_\nu+P{\delta^\mu}_\nu$, Eq. (\ref{eq:Omega-T}) in matrix representation becomes
\begin{equation}
{[\hat \Omega^{-1}]^\mu}_\nu=\frac{1}{2\phi |\hat\Omega|}\left(\begin{array}{cc} w_1^{-1}
 & \vec{0} \\
\vec{0} & w_2^{-1} \hat I_{3\times 3}
\end{array}\right)  \ ,
\end{equation}
where
\begin{eqnarray}
w_1&\equiv& [\phi |\hat\Omega|+V(\phi)+\lambda+\epsilon \kappa^2 \rho]^{-1}  \\
w_2&\equiv& [\phi |\hat\Omega|+V(\phi)+\lambda-\epsilon \kappa^2 P ]^{-1}\ .
\end{eqnarray}
 It is thus easy to see that
\begin{equation}
{\Omega^\mu}_\nu={2\phi|\hat\Omega|}\left(\begin{array}{cc}
w_1 & \vec{0} \\
\vec{0} & w_2\hat I_{3\times 3}
\end{array}\right)  \ .
\end{equation}
The determinant of $\hat \Omega^{-1}$ leads to
\begin{equation}\label{eq:det}
16\phi^4{|\hat\Omega|^3}=1/(w_1w_2^{3}) \ ,
\end{equation}
which establishes an algebraic relation between $|\hat\Omega|$ and the matter.  For the original Born-Infeld theory an explicit relation is easily found in the form $|\hat\Omega|=(\lambda+\epsilon \kappa^2\rho)(\lambda-\epsilon \kappa^2P)^3$.
In general, however, numerical methods will be necessary to find $|\hat\Omega|=|\hat\Omega|(\rho,P)$. Here, for simplicity, we will just consider single perfect fluids with barotropic equation of state, $P/\rho=\omega=$ constant, though more general scenarios can also be considered by just noting that $\rho$ and $P$ represent the total energy density and pressure, respectively. In Figs. \ref{fig:w1b3}, \ref{fig:w1b5} and \ref{fig:wm1b5} we provide graphic representations obtained numerically to illustrate the relation between $|\hat\Omega|$ and $\rho$ for different values of $\omega$ in some particular gravity models.

\subsection{Cosmological models}

Considering, for simplicity, a spatially flat Friedman-Lemaitre-Robertson-Walker line element of the form $ds^2=g_{\mu\nu}dx^\mu dx^\nu=-dt^2+a^2\delta_{ij}dx^i dx^j$, and taking into account the relation
\begin{equation}
t_{\mu\nu}=\phi |\hat\Omega| g_{\mu\alpha}{\Omega^\alpha}_\nu \ ,
\end{equation}
one can verify that the time-time component of the Einstein tensor of the metric $t_{\mu\nu}$ leads to \cite{Olmo_book}
\begin{equation}
3\left(H+\frac{\dot\Delta}{2\Delta}\right)^2=\frac{\kappa^2}{2}\left[\frac{\rho+3P+2(\phi |\hat\Omega|-V-\lambda )/\kappa^2\epsilon}{\phi |\hat\Omega|+V+\lambda+\epsilon \kappa^2 \rho}\right] \ ,
\end{equation}
where $H=\dot a/a$, $\dot \Delta=d \Delta/dt$, and
\begin{equation}
\Delta=\frac{2\phi^2 |\hat\Omega|^{3/2}}{\phi |\hat\Omega|+V+\lambda-\epsilon \kappa^2 P} \ .
\end{equation}
All the above formulae can be applied to a combination of non-interacting perfect fluids by just interpreting $\rho$ and $P$ as the total energy density and pressure of the fluids, i.e., $\rho=\sum_{i=1}^n \rho_i$ and $P=\sum_{i=1}^n P_i$. For fluids  with equation of state $P_i=\omega_i(\rho_i) \rho_i$, the time derivative $\dot \Delta$ can be computed by means of the chain rule,
\begin{equation}
\dot \Delta=\sum_{i=1}^n\left[\frac{\partial \Delta}{\partial  |\hat\Omega|} \frac{\partial  |\hat\Omega|}{\partial \rho_i}+\frac{\partial \Delta}{\partial \rho_i} \right]\dot \rho_i \ ,
\end{equation}
using Eq. (\ref{eq:det}) to obtain the quantities $\frac{\partial  |\hat\Omega|}{\partial \rho_i}$, and using the conservation equations $\dot\rho_i=-3H(1+\omega_i)\rho_i$. The Hubble function thus becomes
\begin{equation}\label{eq:H2final}
\epsilon H^2=\frac{1}{6 \left(1+\frac{\dot\Delta}{2\Delta H}\right)^2}\left[\frac{\epsilon \kappa^2(\rho+3P)+2(\phi |\hat\Omega|-V-\lambda )}{\phi |\hat\Omega|+V+\lambda+\epsilon \kappa^2 \rho}\right] \ ,
\end{equation}
where
\begin{equation}\label{eq:DenomH2}
\frac{\dot\Delta}{2\Delta H}=-\frac{3}{2\Delta}\sum_{i=1}^n\left[\frac{\partial \Delta}{\partial  |\hat\Omega|} \frac{\partial  |\hat\Omega|}{\partial \rho_i}+\frac{\partial \Delta}{\partial \rho_i} \right] (1+\omega_i)\rho_i \ .
\end{equation}

\subsection{Particular examples}

Let us consider the family of theories
\begin{equation}\label{eq:fn}
f(|\hat\Omega|)=|\hat\Omega|^{n}
\end{equation}
For these theories, one can easily verify that the low-energy dynamics smoothly recovers general relativity. In fact, given that $\hat\Omega=\hat I+\epsilon \hat P$, a series expansion in $\epsilon$ yields $|\hat\Omega|\approx 1+\epsilon \text{Tr}[\hat P]+\frac{\epsilon^2}{2}\left( \text{Tr}[\hat P]^2-\text{Tr}[\hat P^2]\right)+O(\epsilon^3)$. Using the fact that $\text{Tr}[\hat P]=R$ and $\text{Tr}[\hat P^2]=R_{\mu\nu}R^{\mu\nu}$, one gets
\begin{equation}
\lim_{\epsilon\to 0} |\hat\Omega|^{n}\approx 1+n\epsilon R+\frac{n\epsilon^2}{2}\left( n R^2-R_{\mu\nu}R^{\mu\nu}\right)+O(\epsilon^3) \ .
\end{equation}
To lowest order, therefore, we find that the action of these theories takes the form
\begin{equation}
\lim_{\epsilon\to 0}S_f=\int d^4 x\sqrt{-g}\left[\frac{(1-\lambda)}{\epsilon\kappa^2}+\frac{n}{\kappa^2}R+O(\epsilon)\right]+S_m \ .
\end{equation}
This shows that with the definitions $\tilde{\kappa}^2\equiv \kappa^2/(2n)$ and $\Lambda=(\lambda-1)/(2n\epsilon)$ the standard Einstein-Palatini theory, namely,
\begin{equation}\label{eq:S_EP}
S_{EP}=\frac{1}{2\tilde{\kappa}^2}\int d^4 x\sqrt{-g}\left[R-2\Lambda\right]+S_m \ ,
\end{equation}
is recovered in the limit $\epsilon \to 0$ for arbitrary $n>0$.
A small departure of $\lambda$ from unity would justify the existence of a cosmological constant. Since the parameter $\epsilon$ has dimensions of a length squared, one can assume that $\epsilon \sim l_P^2$, with $l_P=\sqrt{\hbar G /c^3}$ being the Planck length. The limit $\epsilon \to 0$ thus represents situations in which the curvatures involved are much smaller than the Planck scale. This family of theories, therefore, is in perfect agreement with all currently available observations. Departures only arise at the extremely high curvatures that occur in the very early stages of the universe, which is precisely the regime we are interested in.

\begin{figure}[h]
\begin{center}
\includegraphics[width=0.5\textwidth]{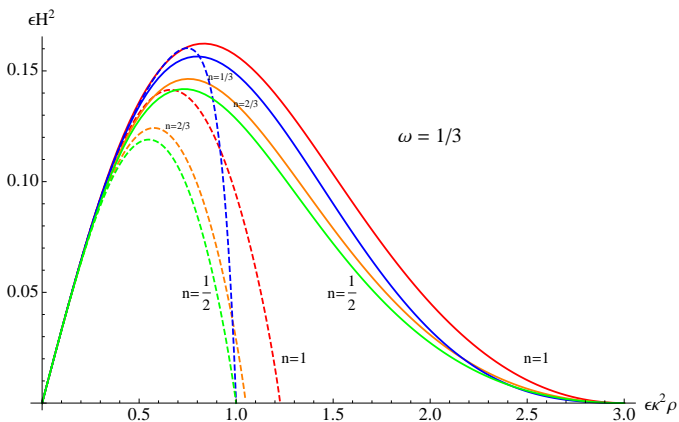}
\caption{Representation of the squared Hubble function for a radiation fluid, $\omega=1/3$, in a family of theories defined as $f(|\hat\Omega|)=|\hat\Omega|^n$ (the Born-Infeld theory corresponds to $n=1/2$ and appears in green), with $\lambda=1$. The dashed lines represent the curves ($-\epsilon \rho, -\epsilon H^2$) and have been plotted in the positive quadrant for convenience. They represent genuine bouncing solutions, whereas the solid ones are unstable solutions with $H^2=0$ and $dH/d\rho=0$ at some high energy density. It is clear that the qualitative behavior of the two families of non-singular solutions persists for small deviations of the parameter $n$. Here we have plotted the cases $n=1/3$ (blue), $n=2/3$ (orange), and $n=1$ (red).   \label{fig:H2a}}
\end{center}
\end{figure}

\begin{figure}[h]
\begin{center}
\includegraphics[width=0.5\textwidth]{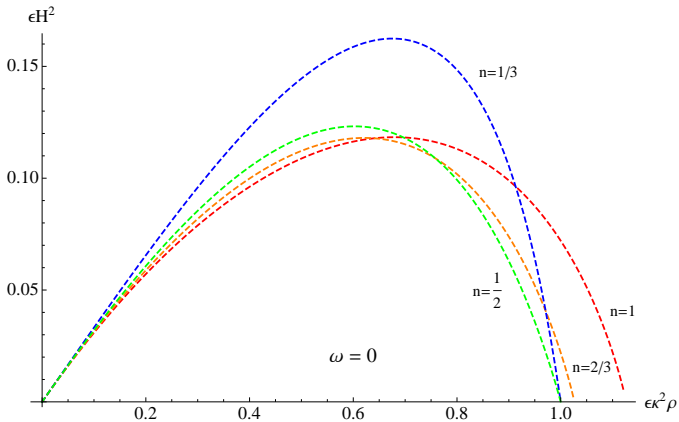}
\caption{Same representation as in Fig. \ref{fig:H2a} but for the case  $\omega=0$. Only those solutions yielding bouncing solutions have been represented, which correspond to the branch $\epsilon<0$ (dashed lines). For $w\leq 0$ no unstable regular solutions exist. \label{fig:H2b}}
\end{center}
\end{figure}

\begin{figure}[h]
\begin{center}
\includegraphics[width=0.5\textwidth]{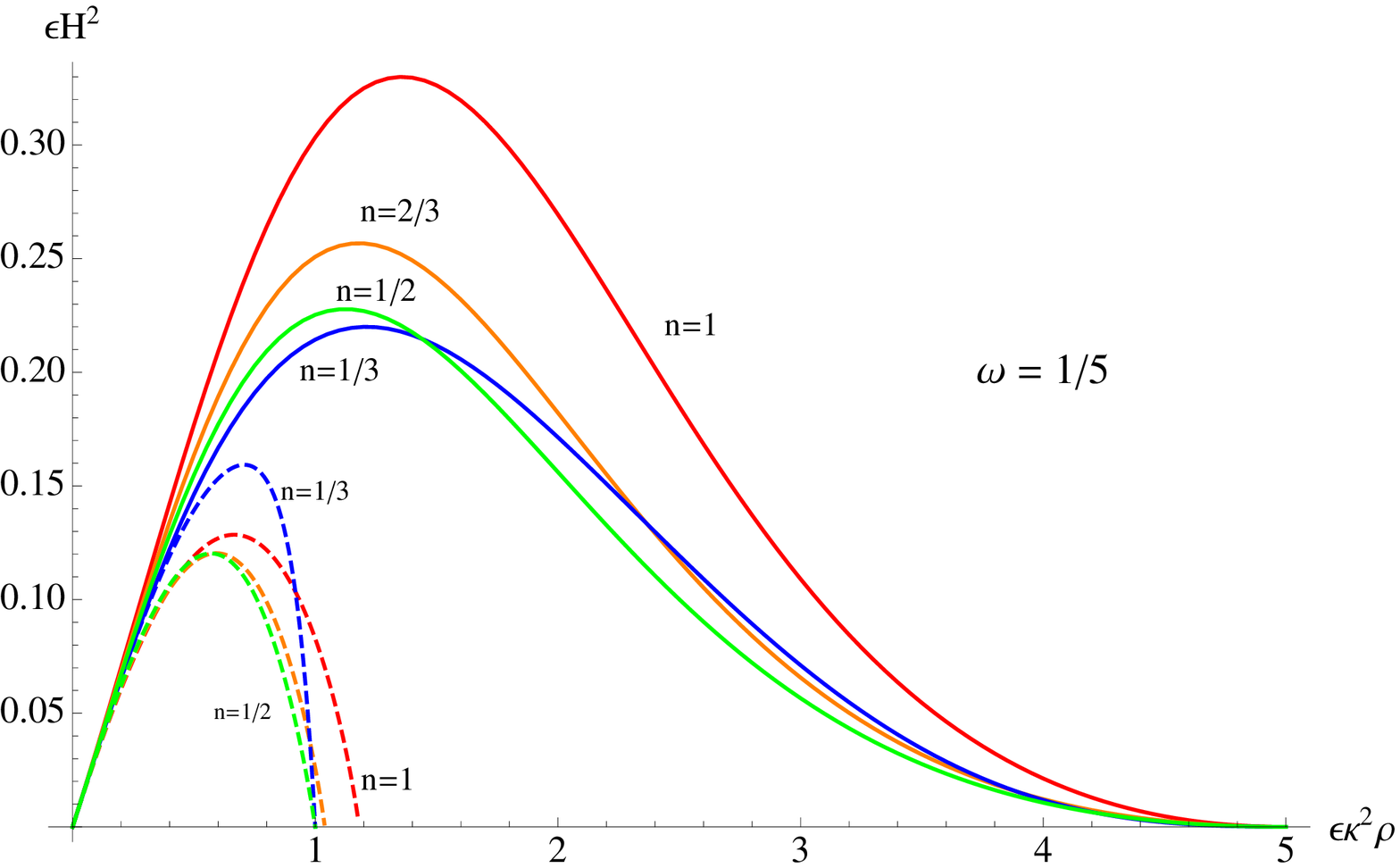}
\caption{Same representation as in Fig. \ref{fig:H2a} but for the case  $\omega=1/5$. \label{fig:H2c}}
\end{center}
\end{figure}

\begin{figure}[h]
\begin{center}
\includegraphics[width=0.5\textwidth]{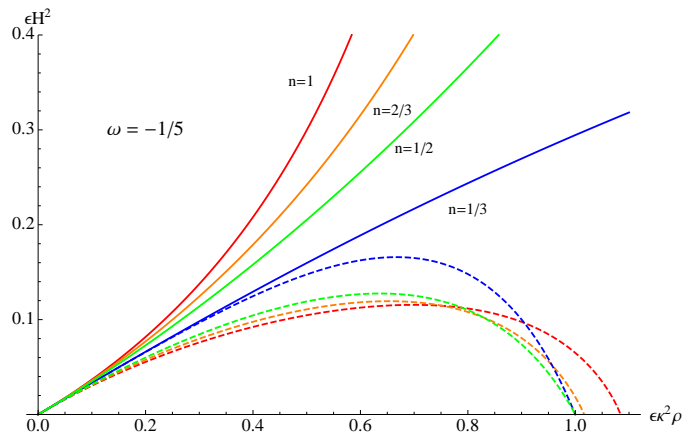}
\caption{Same representation as in Fig. \ref{fig:H2a} but for the case  $\omega=-1/5$. Note that the solid lines are divergent, which indicates that only the case $\epsilon<0$ is able to yield non-singular solutions for arbitrary $\omega$. \label{fig:H2d}}
\end{center}
\end{figure}

For the Lagrangian function (\ref{eq:fn}), we find that $\phi=\frac{df}{d|\hat\Omega|}=n |\hat\Omega|^{n-1}$, which implies $V(\phi)=(n-1) \Omega ^n$ and turns (\ref{eq:det})  into
\begin{eqnarray}
16 n^4 |\hat\Omega|^{4 n-1}&=&\left(\lambda +(2n-1) |\hat\Omega|^n-\epsilon \kappa^2 P\right)^3  \nonumber \\ &\times& \left(\lambda +(2n-1) |\hat\Omega|^n+\epsilon\kappa^2 \rho \right) \ .
\end{eqnarray}
This equation was used to obtain the dependence of $|\hat\Omega|$ on $\rho$ presented in Figs. \ref{fig:w1b3}, \ref{fig:w1b5} and \ref{fig:wm1b5}. With $|\hat\Omega|(\rho)$ known, one can readily obtain the Hubble function using Eqs. (\ref{eq:H2final}) and (\ref{eq:DenomH2}). From Figs. \ref{fig:H2a}, \ref{fig:H2b},  and \ref{fig:H2c} one verifies that the essential features of the original Born-Infeld theory are qualitatively preserved for values of the parameter $n$ not too far from $n=1/2$. In fact, for $\omega>0$ we find two types of non-singular solutions at high energies (for which $H^2$ vanishes at some finite density): i) those corresponding to the $\epsilon \rho>0$ branch (solid curves), and ii) those corresponding to the $\epsilon \rho <0 $ branch (dashed curves). In the Born-Infeld case, the solutions of type i) only exist for $\omega>0$  and are characterized by $H^2\propto (\rho_B-\rho)^2 \propto (a-a_B)^2$ at the maximum density, which represent unstable configurations with a minimum volume at past infinity, where $a(-\infty)=a_B$. Type ii) solutions exist for all $\omega>-1$ and are genuine bouncing solutions with $H^2\propto (\rho_B-\rho) \propto (a-a_B)\propto (t-t_B)^2$. For $n\neq 1/2$ and  $\omega\leq 0$ we only observe this kind of regular solutions, as the branch $\epsilon>0$ only yields divergent universes with a big bang [see Fig. \ref{fig:H2d} for details].

\begin{figure}[h]
\begin{center}
\includegraphics[width=0.5\textwidth]{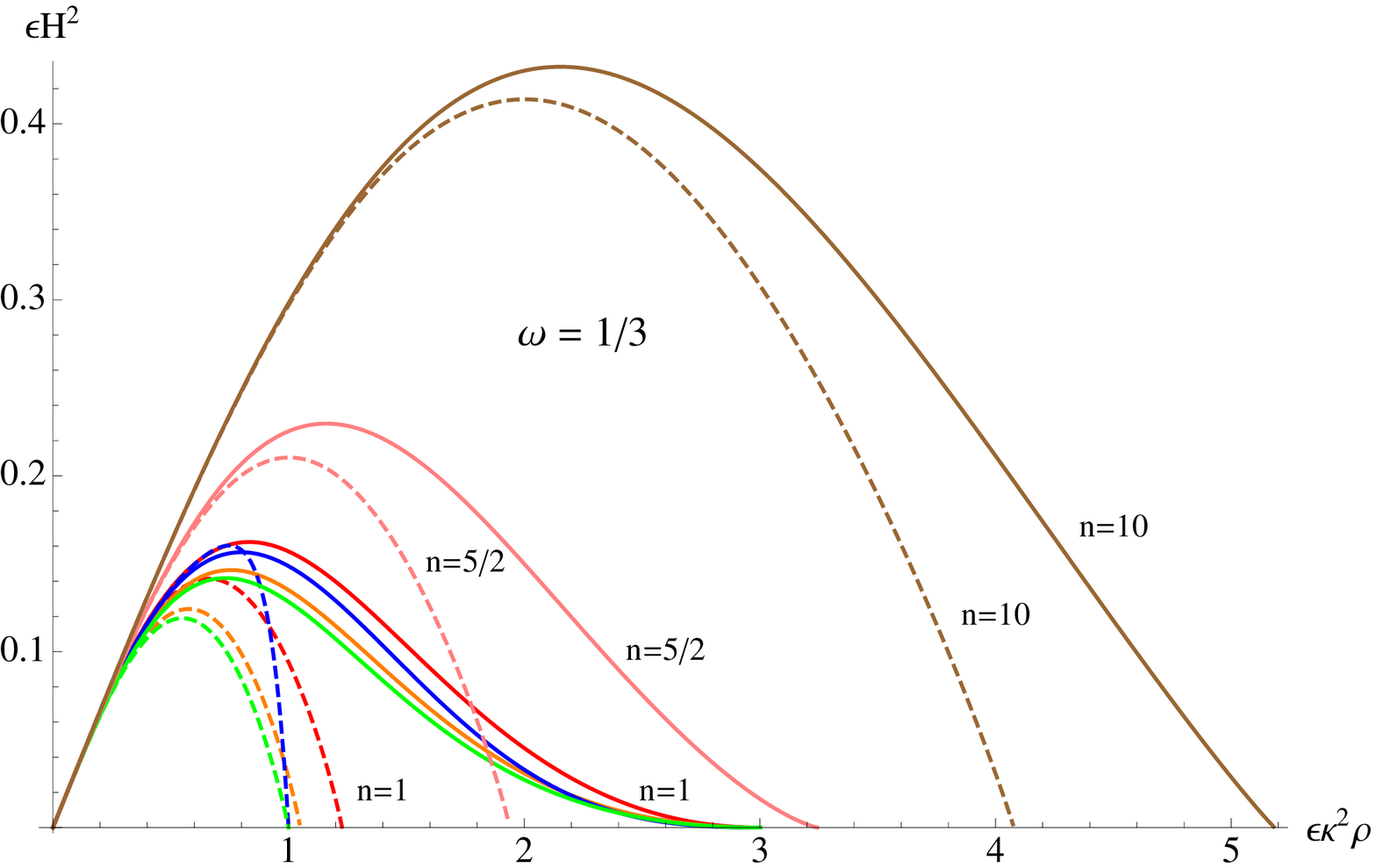}
\caption{Same representation as in Fig. \ref{fig:H2a} but for the case  $\omega=-1/5$. Note that the solid lines for large $n$ hit the horizontal axis forming a non-zero angle, which indicates that they are closer to the type ii)  bouncing solutions of the original Born-Infeld theory than to the unstable type i) solutions with $dH/d\rho =0$ at the maximum density. \label{fig:H2e}}
\end{center}
\end{figure}

In Fig. \ref{fig:H2e}  we can verify that the soft decay ($H^2\to 0$ and $dH/d\rho \to 0$) of the solid curves ($\epsilon>0$) gets modified for large values of $n$. In fact, as $n$ grows from $n=5/2$ to $n=10$ we see that the tendency of $H^2$ is to hit the horizontal axis with a non-zero angle ($dH/d\rho \neq 0$). This indicates that the qualitative properties of these non-singular solutions becomes closer to the bouncing solutions than to the unstable  solutions of smaller $n$'s. We thus conclude that type i) solutions of the Born-Infeld theory are more sensitive to changes in the index $n$ than type ii), which are quite robust. This sensitivity of type i) solutions to changes in the Lagrangian was already observed in \cite{Makarenko:2014lxa} when $R^2$ corrections were included via an $f(R)$ term. In that case, however, the solutions had a tendency to become singular, though other interesting aspects such as the existence of a long inflationary phase after the big bang were also observed. Type ii) solutions were found to be robust even under $f(R)$ corrections as well.

\section{Summary and conclusions}

In this work we have shown that the Born-Infeld theory can be seen as a particular example of a family of $f(|\hat\Omega|)$ theories, where $|\hat\Omega|$ is the determinant of a matrix $\hat \Omega=\hat g^{-1} \hat q$, with $q_{\mu\nu}=g_{\mu\nu}+\epsilon R_{\mu\nu}(\Gamma)$, with the function $f(|\hat\Omega|)$ taking the particular form $f(|\hat\Omega|)=|\hat\Omega|^{1/2}$. We have then studied the field equations of the $f(|\hat\Omega|)$ family of theories, which can be put in Einstein-like form using methods  developed in previous works,  and considered applications to the early-time cosmology to determine whether the avoidance of cosmic singularities is a robust prediction or not. For this purpose we have focused on the class of theories $f(|\hat\Omega|)=|\hat\Omega|^n$ and have shown that their analysis is feasible using elementary numerical methods to solve the algebraic relation between $|\hat\Omega|$ and the energy density $\rho$. The graphic representation of the function $H^2$ as a function of $\rho$ confirms the qualitative robustness of the Born-Infeld predictions. In fact, we have found that the bouncing solutions persist in all the studied models (from $n=1/3$ up to $n=10$). The unstable solutions of the Born-Infeld theory, which represent regular states of minimum volume at past infinity, seem to persist for small values of $n$ but turn into bouncing solutions for larger $n$'s, as is clear from Fig.(\ref{fig:H2e}) for the cases $n=5/2$ and $n=10$.  We thus conclude that the avoidance of cosmic big bang singularities in Born-Infeld and Born-Infeld like theories \cite{Makarenko:2014lxa} is a rather robust phenomenon that does not require fine-tuning in the parameters of the theory, and a generic prediction of Palatini gravities \cite{Olmo_book}. The smooth recovery of the dynamics of GR at lower curvatures [see the discussion above Eq.(\ref{eq:S_EP})] for arbitrary values of $n>0$ is another positive aspect of these theories, as it guarantees their consistency with observations. In this respect, we note that theories of the $f(R)$ type with $f(R)=R^n$ are unable to recover the predictions of general relativity unless $n\sim 1$ \cite{MyReview,Olmo:2005hc}.

As pointed out in the introduction, the Born-Infeld  algorithm allows to improve the behavior of classical theories establishing bounds to certain quantities which otherwise could be divergent, such as the speed of a particle, the electromagnetic field strength, or the curvature invariants of a gravity theory. In the latter case, working in a metric-affine (or Palatini) framework is essential to keep the second-order degree of the equations and to avoid ghosts. The impact of the Born-Infeld high-energy dynamics on black hole structure has been investigated recently finding that curvature singularities can also be avoided in those scenarios in some cases \cite{Olmo:2013gqa}. The results presented here motivate further analyses in that direction to determine the impact that higher values of $n$ could have on the internal structure of black holes. Our construction also permits the consideration of non-minimal matter-geometry couplings with Lagrangians of the form $f(|\hat\Omega|)+g(|\hat\Omega|)\LL_m$, with $\LL_m$ representing the matter Lagrangian. Similar types on non-minimally coupled theories were reviewed in metric formalism in \cite{Nojiri:2010wj} and have been recently studied in Palatini formalism in \cite{Olmo:2014sra}. The cosmology and black hole structure of this type of non-minimal models  will be explored elsewhere.

\section*{Acknowledgments}

S.D.O. is supported in part by MINECO (Spain), project FIS2010-15640 and by grant of Russ. Min. of Education and Science, project TSPU-139.
G.J.O. is supported by MINECO project FIS2011-29813-C02-02, the Consolider Program CPANPHY-1205388, the JAE-doc program of the Spanish Research Council (CSIC), and the i-LINK0780 grant of CSIC. D.R.-G. is supported by CNPq (Brazil agency) through project No. 561069/2010-7 and acknowledges the Departament of Physics of Valencia U. for partial support under project FIS2011-29813-C02-02. G.J.O. and  D.R.-G. also acknowledge funding support of CNPq project No. 301137/2014-5.

\end{document}